\documentclass[nofootinbib,twocolumn,showpacs,preprintnumbers,pre,aps]{revtex4-1}

\usepackage[dvipdfmx]{graphicx}% Include figure files eps, pdf etc.
\usepackage{bm}% bold math
\usepackage{amsmath}
\usepackage{amssymb}
\usepackage{color}
\begin{document}

\title{Odd Microswimmer}%

\author{Kento Yasuda}
\email{Present address: Research Institute for Mathematical Sciences, 
Kyoto University, Kyoto 606-8502, Japan}

\affiliation{
Department of Chemistry, Graduate School of Science,
Tokyo Metropolitan University, Tokyo 192-0397, Japan}

\author{Yuto Hosaka}

\affiliation{
Department of Chemistry, Graduate School of Science,
Tokyo Metropolitan University, Tokyo 192-0397, Japan}

\author{Isamu Sou}

\affiliation{
Department of Chemistry, Graduate School of Science,
Tokyo Metropolitan University, Tokyo 192-0397, Japan}

\author{Shigeyuki Komura}\email{komura@tmu.ac.jp}

\affiliation{
Department of Chemistry, Graduate School of Science,
Tokyo Metropolitan University, Tokyo 192-0397, Japan}

\date{\today}

\begin{abstract}
We propose a model for a thermally driven microswimmer in which three spheres are 
connected by two springs with odd elasticity. 
We demonstrate that the presence of odd elasticity leads to the directional locomotion of the 
stochastic microswimmer.
\end{abstract}
%\pacs{}

%\keywords{Suggested keywords}%Use show keys class option if keyword
\maketitle

Although micromachines such as proteins and enzymes experience the influence of 
strong thermal fluctuations, they often exhibit directional locomotion under nonequilibrium 
conditions~\cite{Yuan21}.
To elucidate this type of phenomena, we previously proposed a thermally driven 
elastic microswimmer consisting of three spheres~\cite{Hosaka17}. 
In this model, the three spheres were assumed to be in equilibrium with independent 
heat baths characterized by different temperatures.

Recently, Scheibner \textit{et al.}\ introduced the concept of ``odd elasticity," which 
can arise from active and nonreciprocal
interactions~\cite{Scheibner20}.
Importantly, the odd part of the elastic constant tensor quantifies the amount of 
work extracted along quasistatic deformation cycles.
In this paper, we propose a novel type of thermally driven microswimmer in which 
the three spheres are connected with springs having not only even elasticity~\cite{Yasuda17}, 
but also odd elasticity~\cite{Scheibner20}.  
We explicitly demonstrate that the proposed stochastic ``odd microswimmer" can exhibit 
a directional locomotion as a result of odd elasticity.
Additionally, we provide a simple physical interpretation of the average velocity 
within the nonequilibrium statistical physics.

\begin{figure}[t]
\begin{center}
\includegraphics[scale=0.4]{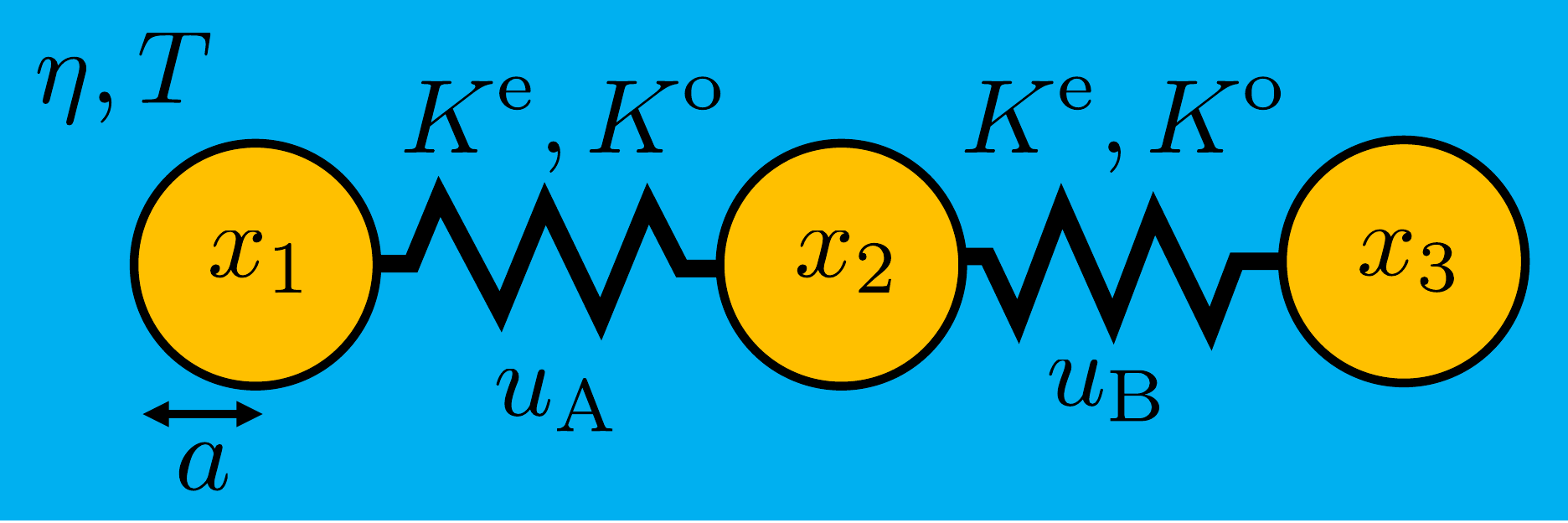}
\end{center}
\caption{(Color online) Odd microswimmer in a fluid with a viscosity $\eta$ and temperature $T$. 
Three spheres of radius $a$ are connected by two springs with a natural length $\ell$.
Each spring has both even elastic constant $K^\mathrm e$ and odd elastic constant 
$K^\mathrm o$. 
The positions of the spheres are denoted as $x_i$ ($i=1,2,3$), and the spring 
extensions with respect to $\ell$ are denoted as $u_\mathrm A$ and $u_\mathrm B$.}
\label{Fig:model}
\end{figure}

Consider a three-sphere microswimmer in which the positions of the three spheres
of radius $a$ are given by $x_i$ ($i=1, 2, 3$) in a one-dimensional coordinate system
(see Fig.~\ref{Fig:model})~\cite{Golestanian08}. 
These three spheres are connected by two springs that exhibit both even and odd 
elasticity.
We denote the two spring extensions as 
$u_\mathrm{A}=x_2-x_1-\ell$ and $u_\mathrm{B}=x_3-x_2-\ell$, where $\ell$ is the natural length.
Then, the forces $F_\mathrm A$ and $F_\mathrm B$ conjugate to $u_\mathrm A$ and 
$u_\mathrm B$, respectively, are given by $F_\alpha=-K_{\alpha\beta}u_\beta$ 
($\alpha, \beta = \mathrm A, \mathrm B$).
For an odd spring, the elastic constant $K_{\alpha\beta}$ is given by~\cite{Scheibner20} 
\begin{align}
K_{\alpha\beta}=K^\mathrm{e}\delta_{\alpha\beta}+K^\mathrm{o}\epsilon_{\alpha\beta}, 
\label{ElasticConstant}
\end{align}
where $K^\mathrm{e}$ and $K^\mathrm{o}$ are the even and odd elastic constants, respectiverly,
in the 2D configuration space spanned by $u_\mathrm{A}$ and $u_\mathrm{B}$
(unlike the real 2D space in Ref.~\cite{Scheibner20},
$\delta_{\alpha\beta}$ is the Kronecker delta, and $\epsilon_{\alpha\beta}$ is the 2D Levi-Civita tensor 
with $\epsilon_{\mathrm{AA}}=\epsilon_{\mathrm{BB}}=0$ and 
$\epsilon_{\mathrm{AB}}=-\epsilon_{\mathrm{BA}}=1$.
The presence of odd elasticity $K^\mathrm{o}$ in Eq.~(\ref{ElasticConstant}) 
reflects the nonreciprocal interaction between the two springs such that $u_\mathrm A$ and $u_\mathrm B$ 
influence each other in a different manner~\cite{Era21}.
The forces $f_i$ acting on each sphere are given by 
$f_1=-F_\mathrm A$, $f_2=F_\mathrm A-F_\mathrm B$, and $f_3=F_\mathrm B$.
These forces satisfy the force-free condition, i.e., $f_1+f_2+f_3=0$.

The odd microswimmer described above is immersed in a fluid with a shear viscosity of $\eta$ 
and temperature $T$. 
Then the equations of motion for each sphere are given by~\cite{Hosaka17,Yasuda17,Golestanian08}
\begin{align}
\dot x_i=M_{ij} f_j+\xi_i,
\label{DynamicEq}
\end{align}
where $\dot x_i=dx_i/dt$ and $M_{ij}$ are the hydrodynamic mobility coefficients~\cite{Golestanian08}
\begin{align}
M_{ij}=\begin{cases}
    1/(6\pi\eta a) & (i=j) \\
    1/(4\pi\eta \, |x_i-x_j|) & (i \ne j)
\end{cases}.
\label{Mobility}
\end{align}
In Eq.~(\ref{DynamicEq}), the Gaussian white-noise sources $\xi_i$ have zero mean 
$\langle\xi_i (t) \rangle=0$, and their correlations satisfy the following 
fluctuation-dissipation theorem:
\begin{align}
\langle\xi_i(t)\xi_j(t')\rangle=2k_\mathrm B T M_{ij}\delta(t-t'),
\label{FDT}
\end{align}
where $k_\mathrm B$ is the Boltzmann constant.

It is convenient to introduce a characteristic time scale $\tau=6\pi\eta a/K^\mathrm e$ 
and the ratio between the two spring constants $\lambda=K^\mathrm o/K^\mathrm e$.
In the following analysis, we assume $u_\mathrm A, u_\mathrm B \ll \ell$ 
and $a \ll \ell$, and focus solely on the leading-order contribution.
The total velocity of the microswimmer is given by $V=(\dot x_1+\dot x_2+\dot x_3)/3$.
After taking the statistical average and using Eqs.~(\ref{ElasticConstant})-(\ref{Mobility}), 
we obtain~\cite{Hosaka17} 
\begin{align}
\langle V\rangle&=\frac{a}{8\ell^2\tau}\left[\langle u_\mathrm B^2\rangle-\langle u_\mathrm A^2\rangle+\lambda\left(3\langle u_\mathrm B^2\rangle+3\langle u_\mathrm A^2\rangle-2\langle u_\mathrm Au_\mathrm B\rangle\right)\right]\nonumber\\
&+\mathcal O[(a/\ell)^2,(u/\ell)^3],
\label{genvelocity}
\end{align}
where we use $\langle u_\mathrm{A}\rangle=\langle u_\mathrm{B}\rangle=0$.

The equal-time correlation functions appearing in Eq.~(\ref{genvelocity}) can be obtained from 
the reduced Langevin equations for $\dot u_\mathrm A=\dot x_2-\dot x_1$ and 
$\dot u_\mathrm B=\dot x_3-\dot x_2$ as 
\begin{align}
\dot u_\alpha=\Gamma_{\alpha\beta}u_\beta+\Xi_\alpha+\mathcal O[a/\ell],
\label{DynamicEqu}
\end{align}
where $\Gamma_{\alpha\beta}$ and $\Xi_\alpha$ are 
\begin{align}
\mathbf{\Gamma}=  -\frac{1}{\tau} \left(
    \begin{array}{ccc}
      2+\lambda & -1+2\lambda \\
      -1-2\lambda & 2-\lambda
    \end{array}
  \right),~~~
\mathbf{\Xi}=  \left(
    \begin{array}{cc}
      \xi_2-\xi_1 \\
      \xi_3-\xi_2
    \end{array}
  \right).
  \label{GammaXi}
\end{align}
Notice that $\Gamma_{\alpha\beta}$ is nonreciprocal, i.e., $\Gamma_\mathrm{AB}\neq \Gamma_\mathrm{BA}$
when $\lambda \neq 0$.
By solving Eq.~(\ref{DynamicEqu}) in the Fourier domain and using Eq.~(\ref{FDT}), we obtain the 
following equal-time correlation functions~\cite{Hosaka17}:
\begin{align}
& \langle u_\mathrm A^2\rangle=\frac{k_\mathrm BT}{K^\mathrm e}\left[1-\frac{\lambda}{2(1+\lambda^2)}\right]+\mathcal O[a/\ell],
\label{uA2}\\
& \langle u_\mathrm B^2\rangle=\frac{k_\mathrm BT}{K^\mathrm e}\left[1+\frac{\lambda}{2(1+\lambda^2)}\right]+\mathcal O[a/\ell],
\label{uB2}\\
& \langle u_\mathrm Au_\mathrm B\rangle=-\frac{k_\mathrm BT}{K^\mathrm e}\frac{\lambda^2}{2(1+\lambda^2)}+\mathcal O[a/\ell].
\label{uAuB}
\end{align}
Here, we neglect the cross-correlations $\langle\xi_i\xi_j\rangle$ with $i\ne j$ because 
they only contribute to higher orders in $a/\ell$.
When $\lambda=0$, the above expressions reduce to 
$\langle u_\mathrm A^2\rangle= \langle u_\mathrm B^2\rangle=k_\mathrm B T/K^\mathrm e$ and 
$\langle u_\mathrm Au_\mathrm B\rangle=0$, reproducing the thermal equilibrium situation.
We have $\langle u_\mathrm A^2\rangle < \langle u_\mathrm B^2\rangle$ when 
$\lambda>0$, because the effective elastic constant of spring A is greater than that of spring B.

By substituting Eqs.~(\ref{uA2})-(\ref{uAuB}) into Eq.~(\ref{genvelocity}), we obtain the average velocity as
\begin{align}
\langle V\rangle=\frac{7ak_\mathrm BT\lambda}{8\ell^2 K^\mathrm e\tau}+\mathcal O[(a/\ell)^2,(u/\ell)^3].
\label{velocity}
\end{align}
Here, $\langle V\rangle$ is proportional to the odd elastic constant $K^\mathrm{o}$ that can take 
either positive or negative value.
Because $\langle V\rangle$ is also proportional to $k_\mathrm BT$, thermal fluctuations are responsible for
the locomotion of the odd microswimmer. 
Therefore, our model provides a novel type of Brownian ratchet.

Next, we discuss the nonequilibrium statistical properties of the odd 
microswimmer~\cite{Sou19,Sou21}.
For the time-dependent probability distribution function $p(u_\mathrm{A},u_\mathrm{B},t)$, 
the Fokker-Planck equation corresponding to Eq.~(\ref{DynamicEqu}) can be written as 
$\dot p=-\partial_\alpha j_\alpha$,
where $\partial_\alpha=\partial/(\partial u_\alpha)$ and $j_\alpha$ is the probability flux given by~\cite{Sou19} 
\begin{align}
j_\alpha=\Gamma_{\alpha\beta}u_\beta p-D_{\alpha\beta} \partial_\beta p.
\label{Flux}
\end{align}
Here, $D_{\alpha\beta}$ is the diffusion matrix 
\begin{align}
\mathbf{D}=  \frac{k_\mathrm BT}{6\pi\eta a}\left(
    \begin{array}{cc}
      2& -1\\
      -1&2
    \end{array}
  \right),
  \label{DM}
\end{align}
which satisfies the relationship $\langle\Xi_\alpha(t)\Xi_\beta(t')\rangle=2D_{\alpha\beta}\delta(t-t')$ 
according to Eq.~(\ref{FDT}).

Owing to the reproductive property of Gaussian distributions, the steady-state probability distribution 
function that satisfies $\dot p=0$ is given by a Gaussian function~\cite{Sou19} 
\begin{align}
p(u_\mathrm{A},u_\mathrm{B})=\frac{1}{2\pi \sqrt{\det \mathbf C}}
\exp\left[-\frac{1}{2}(C^{-1})_{\alpha\beta}u_\alpha u_\beta\right].
\label{PDF}
\end{align}
Here, $C_{\alpha\beta}=\langle u_\alpha u_\beta\rangle$ is the covariance matrix obtained from Eqs.~(\ref{uA2})-(\ref{uAuB}) as 
\begin{align}
\mathbf{C}=\frac{k_\mathrm BT}{K^\mathrm e}\frac{1}{1+\lambda^2}  \left(
    \begin{array}{cc}
      1-\lambda/2+\lambda^2 & -\lambda^2/2 \\
      -\lambda^2/2 & 1+\lambda/2+\lambda^2
    \end{array}
  \right),
  \label{CVM}
\end{align}
and $(C^{-1})_{\alpha\beta}$ is the inverse matrix of $C_{\alpha\beta}$.
For our purposes, we explicitly show that 
\begin{align}
\det \mathbf{C}=\left(\frac{k_\mathrm BT}{K^\mathrm e}\right)^2\frac{4+7\lambda^2+3\lambda^4}{4(1+\lambda^2)^2}.
\label{detCVM}
\end{align}

\begin{figure}[t]
\begin{center}
\includegraphics[scale=0.35]{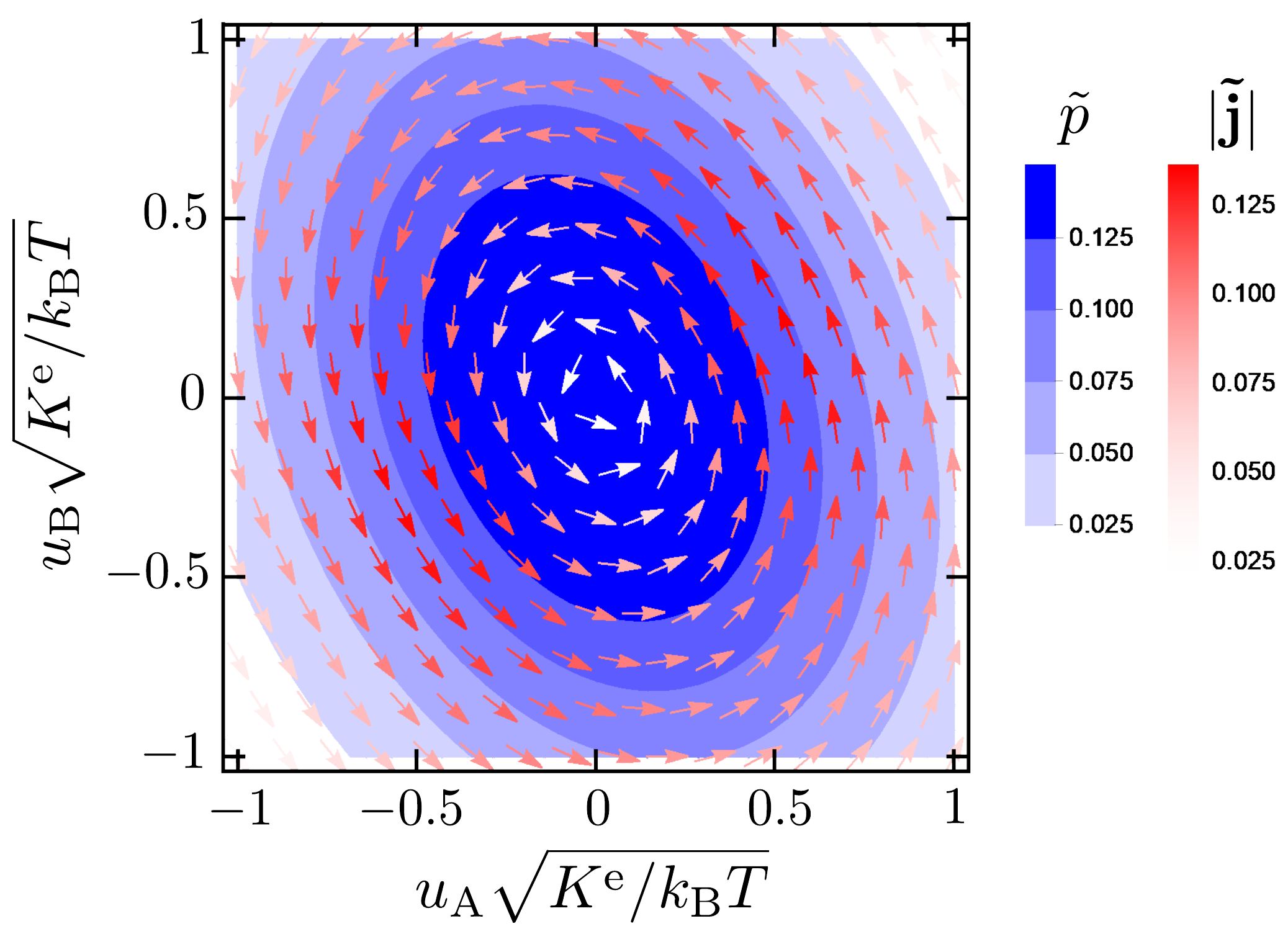}
\end{center}
\caption{(Color online) Steady-state scaled probability distribution function 
$\tilde p=p k_\mathrm{B}T/K^\mathrm{e}$ 
and steady-state scaled probability flux 
$\tilde{\mathbf{j}}=\mathbf{j} \tau\sqrt{k_\mathrm{B}T/K^\mathrm{e}}$ 
(arrows) in the configuration space spanned by 
$u_\mathrm{A}$ and $u_\mathrm{B}$ when $\lambda=K^\mathrm o/K^\mathrm e=1$.
}
\label{Fig:flux}
\end{figure}

In Fig.~\ref{Fig:flux}, we plot the steady-state probability distribution function in Eq.~(\ref{PDF}) and  
corresponding probability flux in Eq.~(\ref{Flux}) when $\lambda=1$.
The probability distribution function is distorted by the negative correlation 
($C_\mathrm{AB}=C_\mathrm{BA}\sim -\lambda^2/2$) between $u_\mathrm{A}$ and $u_\mathrm{B}$.
One can see a counter-clockwise loop of the probability flux.
Such a probability flux becomes clockwise for $\lambda <0$ and vanishes when $\lambda=0$. 
The existence of a probability flux loop indicates that the detailed balance is broken in the  
nonequilibrium steady state.

The steady-state probability flux can be conveniently expressed in terms of a frequency 
matrix $\Omega_{\alpha\beta}$ as $j_\alpha=\Omega_{\alpha\beta} u_\beta p$~\cite{Sou19}.
For the proposed odd microswimmer, the frequency matrix is given by  
\begin{align}
\mathbf{\Omega}=\frac{3\lambda}{\tau(4+3\lambda^2)}  \left(
\begin{array}{ccc}
-\lambda^2 & -2+\lambda-2\lambda^2 \\
2+\lambda+2\lambda^2 & \lambda^2
\end{array}
\right),
\end{align}
which is traceless.
Then, the two eigenvalues of $\Omega_{\alpha\beta}$ are given by 
\begin{align}
\gamma=\pm \mathrm i \frac{3\lambda}{\tau(4+3\lambda^2)} \sqrt{4+7\lambda^2+3\lambda^4}.
\label{eigenvalue}
\end{align}
Because these eigenvalues are purely imaginary, the probability current in the configuration space is rotational.
Comparing Eq.~(\ref{velocity}) with Eqs.~(\ref{detCVM}) and (\ref{eigenvalue}), we obtain the following
simple expression for the average velocity: 
\begin{align}
\vert \langle V\rangle \vert =\frac{7a}{12\ell^2}\sqrt{\det \mathbf C}\,\vert \gamma \vert.
\end{align}
Here, $7a/(12\ell^2)$ is the geometrical factor~\cite{Golestanian08}, 
$\sqrt{\det \mathbf C}\sim k_\mathrm BT/K^\mathrm e$ is the explored area in the configuration 
space, and $\vert \gamma \vert \sim \tau^{-1}$ is the speed of the rotational probability flux~\cite{Sou19}.

Finally, we consider the work that can be extracted when odd elasticity exisits~\cite{Scheibner20}.
For the stochastic odd microswimmer, the average power can be evaluated as 
$\langle\dot W\rangle=-K_{\alpha\beta}\langle\dot u_\alpha u_\beta \rangle$, where 
$W=\int du_\alpha \,F_\alpha$.
From Eq.~(\ref{DynamicEqu}), we obtain 
$\langle \dot u_\mathrm Au_\mathrm B\rangle=-\langle \dot u_\mathrm Bu_\mathrm A\rangle
=-3k_\mathrm BT\lambda/(2K^\mathrm e\tau)$ and 
$\langle \dot u_\mathrm Au_\mathrm A\rangle=\langle \dot u_\mathrm Bu_\mathrm B\rangle=0$.
By using these results, we can estimate the power of the odd microswimmer as 
$\langle\dot W\rangle=3k_\mathrm BT\lambda^2/\tau$.
We have confirmed that this power coincides with the average entropy production rate obtained 
by the expression 
$\langle\dot \sigma\rangle=-\mathrm{Tr}\,[\mathbf{\Gamma}
(\mathbf{\Gamma}\mathbf{C}\mathbf{D}^{-1}+\mathbf{I})]$~\cite{Sou21}, where 
$\mathbf{I}$ is the identity matrix.
Therefore, all the extracted work due to odd elasticity is converted into the entropy production. 
It is also useful to note that the average velocity can be alternatively written as 
$\langle V\rangle=7a/(12\ell^2)\langle \dot u_\mathrm B u_\mathrm A\rangle$.

%\begin{acknowledgment}
K.Y.\ and Y.H.\ acknowledge support by a Grant-in-Aid for JSPS Fellows (Grants No.\ 18J21231
and No.\ 19J20271) from the JSPS.
S.K.\ acknowledges support by a Grant-in-Aid for Scientific Research (C) (Grants No.\ 18K03567 and
No.\ 19K03765) from the JSPS, 
and support by a Grant-in-Aid for Scientific Research on Innovative Areas
``Information Physics of Living Matters'' (Grant No.\ 20H05538) from the MEXT of Japan.
%\end{acknowledgment}

%%%%%%%%%%%%%%%


\begin{thebibliography}{99}
%%%%%%%%%%%%%%%

\bibitem{Yuan21}
H. Yuan, X. Liu, L. Wang, and X. Ma, 
Bioactive Materials \textbf{6}, 1727 (2021).

\bibitem{Hosaka17}
Y. Hosaka, K. Yasuda, I. Sou, R. Okamoto, and S. Komura, 
J. Phys. Soc. Jpn. \textbf{86}, 113801 (2017).

\bibitem{Scheibner20}
C. Scheibner, A. Souslov, D. Banerjee, P. Sur\'{o}wka, W. T. M. Irvine, and V. Vitelli, 
Nat. Phys. \textbf{16}, 475 (2020).

\bibitem{Yasuda17}
K. Yasuda, Y. Hosaka, M. Kuroda, R. Okamoto, and S. Komura, 
J. Phys. Soc. Jpn. \textbf{86}, 093801 (2017). 

\bibitem{Golestanian08}
R. Golestanian and A. Ajdari, 
Phys. Rev. E \textbf{77}, 036308 (2008).

\bibitem{Era21}
K. Era, Y. Koyano, Y. Hosaka, K. Yasuda, H. Kitahata, and S. Komura,
EPL \textbf{133}, 34001 (2021).

\bibitem{Sou19}
I. Sou, Y. Hosaka, K. Yasuda, and S. Komura, 
Phys. Rev. E \textbf{100}, 022607 (2019).

\bibitem{Sou21}
I. Sou, Y. Hosaka, K. Yasuda, and S. Komura, 
Physica A \textbf{562}, 125277 (2021).

\end{thebibliography}
\end{document}